# Sequencing seismograms: A panoptic view of scattering in the core-mantle boundary region

**Authors:** D. Kim[1*], V. Lekić[1], B. Ménard[2], D. Baron[3], M. Taghizadeh-Popp[2]

**Affiliations:**

[1]Department of Geology, University of Maryland, College Park, MD 20742, USA.

[2]Department of Physics and Astronomy, Johns Hopkins University, Baltimore, MD 21218, USA.

[3]School of Physics and Astronomy, Tel-Aviv University, Tel Aviv 69978, Israel.

*Corresponding author. Email: dk696@cornell.edu

## Abstract

Scattering of seismic waves can reveal subsurface structures, but usually in a piecemeal way focused on specific target areas. We used a manifold learning algorithm called "the Sequencer" to simultaneously analyze thousands of seismograms of waves diffracting along the core-mantle boundary and obtain a panoptic view of scattering across the Pacific region. In nearly half of the diffracting waveforms, we detected seismic waves scattered by three-dimensional structures near the core-mantle boundary. The prevalence of these scattered arrivals shows that the region hosts pervasive lateral heterogeneity. Our analysis revealed loud signals due to a plume root beneath Hawaii and a previously unrecognized ultralow-velocity zone beneath the Marquesas islands. These observations illustrate how approaches flexible enough to detect robust patterns with little-to-no user supervision can reveal distinctive insights into the deep Earth.

## One Sentence Summary

Sequencing seismograms reveals an ultralow-velocity zone beneath Marquesas, and pervasive scattering near the core-mantle boundary, strongest from a plume root under Hawaii.

## Main Text

Seismic networks record millions of earthquake waveforms every year. The timing, amplitude, polarization and frequency of their constituent arrivals contain precious information about seismic sources and deep Earth structures. For example, unexpected arrivals can reveal waves scattered or multi-pathed by heterogeneities within Earth, down to the core-mantle boundary (CMB). Mapping heterogeneities in the CMB region is important for understanding the fate of subducted slabs, the origin of hot spot volcanism, and the nature of primitive geochemical reservoirs (*1*). This mapping can be accomplished by identifying and interpreting scattered arrivals, which is challenging because it requires distinguishing seismic fluctuations from noise and contextualizing arrival amplitudes and timing in a regime where models are not currently available.

Traditionally, these challenges are overcome by focusing on a specific target area. This allows leveraging geometric arrangements to identify robust signals and aid in their interpretation. A standard procedure is to arrange waveforms by epicentral distance or azimuth to reveal trends in arrivals not predicted by models of the interior. In the Earth's CMB region, robustly identifying scattered waves has led to various discoveries, such as the D″ discontinuity (*2*), ultralow-velocity zones (ULVZs) (*3*), mega-ULVZs (*4*), and abrupt variations in wave speed across the boundaries of large low shear velocity provinces (LLSVPs) that imply compositional heterogeneity (*5*). In all cases, interpreting seismic waveforms would be difficult in isolation, without the context provided by seismic waves on nearby paths. Therefore, previous approaches have limited utility in poorly-sampled regions and do not make use of the full statistical power of waveform datasets that span geographically diverse paths.

We conducted a large-scale, systematic search for seismic waves scattered by heterogeneity near the CMB across the Pacific basin. We focused on shear waves diffracting along the CMB (Sdiff), because they sample large areas at the base of the mantle. Waves scattered by heterogeneity in this region arrive after the main Sdiff phase, so the timing and amplitude of these Sdiff "postcursors" can constrain the location and nature of structures producing the scattering (*4, 6-9*). Because energy reflected from the surface can complicate the identification of Sdiff postcursors, we restricted our attention to waveforms from deep earthquakes (*10*). This yielded a dataset of ~6000 transverse-component waveforms aligned on the main Sdiff arrival and deconvolved with synthetics computed for a one-dimensional (1D) preliminary reference Earth model (PREM) (*11*). When plotted by distance between the source and the receiver (Fig. 1A), postcursors cannot readily be identified. Ordering by azimuth only makes sense locally and is not useful when working with data spanning large geographic areas.

We made use of an unsupervised graph-based manifold learning algorithm, "the Sequencer" (*10*), that orders objects to minimize dissimilarities between neighbors as well as globally across the entire sequence. In our case, the objects are Sdiff waveforms and dissimilarity is given by the Wasserstein metric (also known as the earth mover's distance) between them (*10*). This approach can be used to reveal the main trend present in a dataset without requiring any model at all; it has been used in astronomy and has already led to the discovery of a trend relating the mass of supermassive black holes to the properties of their host galaxies (*12*). We used the Sequencer to optimally order our collection of Sdiff waveforms across the Pacific basin (Fig. 1B). We identified postcursors in >40% of waveforms, which indicates that postcursors are far more common than previously thought. Radial gradients in velocities cannot on their own explain the postcursors, regardless of ordering (figs. S1, B and D, and S2).

We then explored the geographic distribution of heterogeneities giving rise to the postcursors by binning the fraction of waveforms with postcursors in 1° radius bins (Fig. 1C). We found that heterogeneities large enough to scatter shear waves with periods >15s are pervasive across the Pacific basin. In most locations, a substantial fraction (>0.3) of waves show postcursors (Fig. 1C, cyan). Postcursors are typically absent (Fig. 1C, peach) on paths that do not cross LLSVP boundaries and instead are confined either within the boundaries or well outside of them. Waveforms with and without postcursors seem to coexist at sub-wavelength scales (~160 km) in most locations (Fig. 1C). Many of the postcursors in the northern and western Pacific have large delay times (fig. S3B), indicating that they originate from distant scatterers and may travel through complex structures. These areas are suspected to host a partial melt created by paleo-slab from northwestern Pacific subduction zones (*13*) and a group of small ULVZ patches, along with slab debris beneath the northeastern boundary of the mid-Pacific LLSVP (*14*).

Although large amplitude postcursors appear to be associated with a few localities south of the Aleutians (Fig. 1D, R3), bootstrap error estimates show that these signals are not statistically significant (fig. S4). Notably, nearly all Sdiff waves sampling near Hawaii and the Marquesas Islands shows postcursors (Fig. 1C, pink). The amplitude of postcursors (with respect to the main Sdiff) (Fig. 1D) in the Hawaiian region appears to be three times larger than the typical amplitude found throughout the Pacific basin (Fig. 1D).

When contextualized across all available data using the Sequencer, the region to the northwest of Hawaii stands out in terms of the prevalence of postcursors, their amplitude, and the spatial extent of the area associated with high-amplitude postcursors (Fig. 1, C and D). Therefore, we zoomed in on this region and performed a similar analysis of waveforms with turning points within 20° of Hawaii (Fig. 2A). The nature of such high-amplitude signals is discussed in the methods and materials section of the supplementary materials (*10*). By plotting the waveforms in the optimal order determined by the Sequencer, we readily identified a subpopulation featuring strong postcursors (Fig. 2B). Because of the limited geographic area, this population can also be identified by sorting the waveforms according to azimuth, with delay times increasing up to 40s (fig. S5A). However, when visualized in the order determined by the Sequencer, the moveout of the postcursors is clarified, and coherent geographic patterns can be mapped.

By plotting each waveform's position in the sequence at the midpoint of its diffracting path, we found that waveforms with postcursors appear to the northwest of Hawaii, whereas those without appear predominantly to the southeast. Moreover, the order identified by the Sequencer reveals a distinctive spatial pattern (Fig. 2E), with waveforms appearing late in the sequence flanking a region in which mid-sequence waveforms cluster. This coherent pattern is reflected in both delay time and amplitude of the postcursors observed northwest of Hawaii (Fig. 2, C to E) and does not follow a simple azimuthal trend. Postcursor delay times gradually increase towards the center of the cluster, whereas large amplitude postcursors (above ~0.5 with respect to the main Sdiff phase) are found mostly southward, near the northern edge of Pacific LLSVPs (Fig. 2E). The small number of weak postcursors detected to the southeast cluster does not show coherent geographic trends, suggesting that they were not produced by a structure of the type located beneath Hawaii.

We also observed an anti-correlation between postcursor delay time and amplitude in the region (Fig. 2, C and D), consistent with expectations for a localized wavespeed anomaly. This observation motivated us to use the presence of such a correlation as a detector of localized structures. Thus, at each location across the Pacific basin, we estimated the slope and amplitude of the linear fit to the delay time and log amplitude of postcursors with diffracting paths located within 5° of the location. As expected, we found steep negative slopes in the vicinity of Hawaii, across a region that is substantially larger than anywhere else in the Pacific basin (Fig. 3A). In addition, we also detected a similar signature, with a slightly gentler slope, close to the Marquesas (Fig. 3A), indicating a previously unidentified localized wavespeed anomaly. Near both hot spots, the negative slopes are significant at a 99% level of confidence (Fig. 3B). Other locations with possible detections of localized wavespeed anomalies are in the vicinity of Alaska, Kamchatka, and along the northern edge of the Pacific LLSVP, but none are as clear as those near the Marquesas.

After using the Sequencer to reveal the leading trend in our waveform dataset and map the presence of heterogeneity in the Pacific region, we investigated the physical origin of the postcursor signals through waveform modeling. We carried out a systematic suite of wave propagation simulations through candidate structures based on known features of the CMB region (*10*). Seismically imaged structures near the CMB span a wide range of sizes, from ~5000-km LLSVPs at one end (*15, 16*) to ~100-km ULVZs on the other (*3, 17*). So-called "mesoscale" structures have also been documented, including the Perm (*18*) and Kamchatka (*19*) anomalies, as well as unusually large ULVZs (*4, 6, 20*). Therefore, we explored two types of candidate structures: (i) cylindrical regions of reduced or elevated shear wave velocity ($Vs$) with dimensions reminiscent of ULVZs and (ii) undulating boundaries of large low-$V_S$ regions reminiscent of LLSVPs (Fig. 3). We explored the effects of the lateral abruptness of velocity changes across the boundaries of both types of structures.

Our waveform simulations confirmed that postcursor log amplitude decays linearly with delay time up to 20° away from a low-velocity cylinder (Fig. 4A, inset). The slope of this decay (Fig. 3) is controlled by geometric spreading and seismic attenuation, while cylinder height and width and the magnitude of its velocity reduction determine the amplitude. A high velocity cylinder produces postcursors that are four to five times weaker, with similar log amplitude decay (Fig. 4A). Differences in effects of fast and slow anomalies are discussed in the materials and methods (*10*). Tradeoffs among these physical parameters make it impossible to uniquely map slope and zero-delay amplitude to anomaly size, shape, and $V_S$ reduction (Fig. 4A and fig. S6). Nevertheless, we found that a published model for the Hawaii ULVZ (*4, 7, 8*) (fig. S7) with a 20% reduction in $V_S$ [height (H) = 25 km] (Fig. 4A, dashed red lines) matched the amplitude and delay time of the Marquesas postcursors. This indicates the presence of a mega-ULVZ beneath the Marquesas, although our dataset does not constrain its precise location and characteristics.

However, this model fails to reproduce the large postcursor amplitudes we observe near Hawaii. Instead, a narrower but taller (H = 600 km) low-velocity body representing a deeply rooted plume conduit can match the amplitude and delay time of Hawaii postcursors, as does a 50-km-tall cylinder with more gradual boundaries (H = 50 km, with smooth edges) (Fig. 4A). Waveforms computed in the plume model for an earthquake which samples the Hawaii region agree well with observations ($M_B$ in fig. S8 and S9). Such a narrow (< 500-km \-wide) mantle plume is not resolvable by travel-time tomography (*21*) but should become visible in full waveform inversions (*22*) incorporating the postcursor waveforms. Although more complicated models cannot be ruled out, such as a 100-km-tall extremely low $V_S$ (25% reduction) structure embedded in the northern edge of Pacific LLSVPs (*7*), we prefer the plume root structure to explain the amplitudes and waveforms of Hawaii postcursors.

Away from Hawaii and the Marquesas, postcursor amplitude (Fig. 4B, black) is constant with delay time and significantly greater (fig. S10) than the largest post-Sdiff signal in waveforms that do not show a postcursor when sequenced (Fig. 4B, gray). Postcursor amplitudes cannot be attributed to deconvolution artifacts (*10*) (fig. S10). The spatial pattern of postcursor amplitude variations has a characteristic length scale of ~3000 km (fig. S11), which is similar to the Fresnel width for a scattered arrival with a delay time of 30 to 50s. We ruled out that the structure beneath Hawaii alone produces the postcursors observed across the northern Pacific, because our wavefield simulations indicate that regardless of height, this structure cannot produce postcursors of sufficiently large amplitude on paths >20° away from its center (fig. S8). Rather, two scenarios can plausibly explain postcursors identified by the Sequencer across the northern Pacific: (i) scattering from multiple smaller anomalies distributed across the region and (ii) scattering or multipathing across sufficiently laterally abrupt boundaries of the LLSVPs.

Multiple anomalies distributed across the Pacific could have complex geometries or comprise multiple ULVZs with various geometries and sizes (23). A conglomerate of individual ULVZs that are smaller than our wavelength and ubiquitous across the region may produce the pervasive postcursors we observed. Whether these anomalies involve partial melting (24) or compositionally distinct, low $V_S$ materials (25) should be further examined using various body-wave phases at higher frequencies (26).

Alternatively, sharp edges and complexities associated with the LLSVP (27-28) can broaden shear wave pulses (Fig. 1B) that propagate nearby (29) or result in multipathing (9) and scattering that can produce postcursors. As demonstrated by our wavefield simulations, the postcursors resulting from waves approaching the edges of such a large-scale structure head-on (Fig. 4B, light blue) generally produce weaker-than-observed postcursors. On the other hand, postcursors generated from the waves that transmit obliquely across the same 5% impedance contrast (Fig. 4B, lavender) better reproduce observed amplitudes. The effects of the smoothness of those edges (i.e. sharpness of the velocity along the boundary) were negligible. To test postcursor generation with a more realistic model, we have sharpened the bottom 600 km of the SEMUCB-WM1 tomographic model (30), to introduce sharper edges for the LLSVPs ($M_C$ in fig. S8B). Synthetic waveforms produced by this sharpened model provide a moderately-improved fit to the postcursors observed outside the Hawaii region (fig. S9) but could not generate the high amplitude postcursors near Hawaii (Fig. 4A and fig. S8).

Exploring a large dataset with the Sequencer enables a data-driven analysis of seismic waveforms without any prior expectations. We anticipate this approach to be useful for many types of datasets beyond seismograms. Often, observed phenomena are driven by a leading effect or parameter. In such cases, there should exist a 1D manifold representing this underlying trend, even if it exhibits a complex behavior. Manifold learning techniques such as the Sequencer can reveal this leading trend from complex data, which is especially useful in the absence of theoretical guidance. When the leading trend is already known, it can be removed before sequencing, as we did by deconvolving the 1D Earth response from the seismograms. In our case, the trends identified by the Sequencer end up being surprisingly simple. They could also be obtained by plotting the delay time and amplitude of a Gaussian fit to the postcursors. However, this simplicity was not obvious before "sequencing" the dataset. In other words, once one knows what to look for, revealing the trends presented in this work is straightforward and does not require the Sequencer algorithm.

Incorporating higher frequency waveforms would enable us to constrain structures of even smaller scale. These waveforms are harder to interpret but should not pose a challenge to our manifold learning approach. Our wavefield simulations indicate that, given their frequency content (15 to 100s), our data cannot resolve velocity variations at scales smaller than ~50 km. Although geometry of seismic illumination to the lowermost mantle will remain more or less the same, Sequencer-based approaches for identifying trends and anomalous signals in portions of seismograms that host various seismic phases will advance our understanding of deep Earth structures.

Our systematic postcursor search in the Pacific basin, interpreted with insights from waveform simulations, reveals that heterogeneity capable of producing postcursors is widespread in the CMB region. We found that strong postcursors exhibit an anti-correlation between delay time and log amplitude that can be detected locally. The two strongest postcursor signals in the Pacific are related to the Marquesas and a broad region near Hawaii. The previously unknown localized anomaly near the Marquesas has dimensions similar to that of a mega-ULVZ (*4*). The anomaly beneath Hawaii is unique in the Pacific basin, producing postcursors that can only be explained by a structure either substantially larger than a mega-ULVZ or by a plume conduit. Weaker postcursors are observed throughout the Pacific basin and do not exhibit a correlation between delay time and amplitude; they are best explained by scattering from laterally-abrupt edges of LLSVPs (5% $V_S$ contrast). The origins of and relationships among these CMB region structures remain controversial (*31*). Nevertheless, recent analyses suggest that LLSVPs may host relatively undegassed geochemical reservoirs (*32*). Mega-ULVZs, on the other hand, must involve either exotic compositions or reflect the presence of partial melt (*24, 33*), and they have been proposed to host primitive geochemical signatures predating the moon-forming impact (*34*). Therefore, our discovery of a mega-ULVZ beneath the Marquesas offers a test of this hypothesis.

# References

1. A. M. Dziewondski, B. Romanowicz, Deep Earth Seismology: An introduction and overview. *Treatise on Geophysics*, **1**, 1-28 (2015).

2. T. Lay, Q. Williams, E. J. Garnero, The core–mantle boundary layer and deep Earth dynamics. *Nature*, 392(6675), 461 (1998).

3. E. J. Garnero, S. P. Grand, D. V. Helmberger, Low P-wave velocity at the base of the mantle. *Geophysical Research Letters*, **20**(17), 1843-1846 (1993).

4. S. Cottaarr, B. Romanowicz, An unusually large ULVZ at the base of the mantle near Hawaii. *Earth and Planetary Science Letters*, **355**, 213-222 (2012).

5. L. Wen, P. Silver, D. James, R. Kuehnel, Seismic evidence for a thermo-chemical boundary at the base of the Earth's mantle. *Earth and Planetary Science Letters*, **189**(3-4), 141-153 (2001).

6. K. Yuan, B. Romanowicz, Seismic evidence for partial melting at the root of major hot spot plumes. *Science*, **357**(6349), 393-397 (2017).

7. A. To, Y. Fukao, S. Tsuboi, Evidence for a thick and localized ultra low shear velocity zone at the base of the mantle beneath the central Pacific. *Physics of the Earth and Planetary Interiors*, **184**(3-4), 119-133 (2011).

8. A. To, Y. Capdeville, B. Romanowicz, Anomalously low amplitude of S waves produced by the 3D structures in the lower mantle. *Physics of the Earth and Planetary Interiors*, **256**, 26-36 (2016).

9. A. To, B. Romanowicz, Y. Capdeville, N. Takeuchi, 3D effects of sharp boundaries at the borders of the African and Pacific Superplumes: Observation and modeling. *Earth and Planetary Science Letters*, **233**(1-2), 137-153 (2005).

10. Materials and methods are available as supplementary materials.

11. A. M. Dziewonski, D. L. Anderson, Preliminary reference Earth model. *Physics of the earth and planetary interiors*, **25**(4), 297-356 (1981).

12. D. Baron, B. Ménard, Black hole mass estimation for Active Galactic Nuclei from a new angle. *Monthly Notices of the Royal Astronomical Society*, **487**(3), 3404-3418 (2019).

13. Y. Xu, K. D. Koper, Detection of a ULVZ at the base of the mantle beneath the northwest Pacific. *Geophysical Research Letters*, **36**(17) (2009).

14. D. Sun, D. Helmberger, V. H. Lai, M. Gurnis, J. M. Jackson, H. Y. Yang, Slab control on the northeasterneEdge of the Mid-  Pacific LLSVP near Hawaii. *Geophysical Research Letters*, **46**(6), 3142-3152 (2019).

15. S. Cottaar, V. Lekic, Morphology of seismically slow lower-mantle structures. *Geophysical Supplements to the Monthly Notices of the Royal Astronomical Society*, **207**(2), 1122-1136 (2016).

16. E. J. Garnero, A. K. McNamara, S. H. Shim, Continent-sized anomalous zones with low seismic velocity at the base of Earth's mantle. *Nature Geoscience*, **9**(7), 481-489 (2016).

17. S. Yu, E. J. Garnero, Ultralow velocity zone locations: A global assessment. *Geochemistry, Geophysics, Geosystems*, **19**(2), 396-414 (2018).

18. V. Lekic, S. Cottaar, A. Dziewonski, B. Romanowicz, Cluster analysis of global lower mantle tomography: A new class of structure and implications for chemical heterogeneity. *Earth and Planetary Science Letters*, **357**, 68-77 (2012).

19. Y. He, L. Wen, T. Zheng, Seismic evidence for an 850 km thick low-velocity structure in the Earth's lowermost mantle beneath Kamchatka, *Geophysical Research Letters*, **41**, 7073–7079 (2014).

20. M. S. Thorne, E. J. Garnero, G. Jahnke, H. Igel, A. K. McNamara, Mega ultra low velocity zone and mantle flow. *Earth and Planetary Science Letters*, **364**, 59-67 (2013).

21. R. Maguire, J. Ritsema, M. Bonnin, P. E. van Keken, S. Goes, Evaluating the resolution of deep mantle plumes in teleseismic traveltime tomography. *Journal of Geophysical Research: Solid Earth*, **123**(1), 384-400 (2018).

22. S. W. French, B. Romanowicz, Broad plumes rooted at the base of the Earth's mantle beneath major hotspots. *Nature*, **525**(7567), 95 (2015).

23. C. Zhao, E. J. Garnero, M. Li, A. McNamara, S. Yu, Intermittent and lateral varying ULVZ structure at the northeastern margin of the Pacific LLSVP. *Journal of Geophysical Research: Solid Earth*, **122**(2), 1198-1220 (2017).

24. Q. Williams, E. J. Garnero, Seismic evidence for partial melt at the base of Earth's mantle. *Science*, **273**(5281), 1528-1530 (1996).

25. W. L. Mao, H. K. Mao, W. Sturhahn, J. Zhao, V. B. Prakapenka, Y. Meng, J. Shu, Y. Fei, R. J. Hemley,  Iron-rich post-perovskite and the origin of ultralow-velocity zones. *Science*, **312**(5773), 564-565 (2006).

26. T. Lay, Deep Earth Structure – Lower Mantle and D". *Treatise on Geophysics*, **1**, 619-654 (2015).

27. S. Ni, E. Tan, M. Gurnis, D. Helmberger, Sharp sides to the African superplume. *Science*, **296**(5574), 1850-1852 (2002).

28. T. Lay, J. Hernlund, E. J. Garnero, M. S. Thorne, A post-perovskite lens and D"heat flux beneath the central Pacific. *Science*, **314**(5803), 1272-1276 (2006).

29. C. Zhao, E. J. Garnero, A. K. McNamara, N. Schmerr, R. W. Carlson, Seismic evidence for a chemically distinct thermochemical reservoir in Earth's deep mantle beneath Hawaii. *Earth and Planetary Science Letters*, **426**, 143-153 (2015).

30. S. W. French, B. A. Romanowicz, Whole-mantle radially anisotropic shear velocity structure from spectral-element waveform tomography. Geophysical Journal International, 199(3), pp.1303-1327. (2014).

31. A. K. McNamara, A review of large low shear velocity provinces and ultra low velocity zones. *Tectonophysics*, **760**, 199-220 (2019).

32. C. D. Williams, S. Mukhopadhyay, M. L. Rudolph, B. Romanowicz, Primitive Helium is Sourced from Seismically Slow Regions in the Lowermost Mantle. *Geochemistry, Geophysics, Geosystems,* **20**(8), 4130-4145 (2019).

33. J. K. Wicks, J. M. Jackson, W. Sturhahn, Very low sound velocities in iron-rich (Mg,Fe) O: Implications for the core-mantle boundary region. *Geophysical Research Letters*, **37**(15) (2010).

34. A. Mundl, M. Touboul, M. G. Jackson, J. M. Day, M. D. Kurz, V. Lekic, R. T. Helz, R. J. Walker, Tungsten-182 heterogeneity in modern ocean island basalts. *Science*, **356**(6333), 66-69 (2017).

35. B. Steinberger, Plumes in a convecting mantle: Models and observations for individual hotspots. *Journal of Geophysical Research: Solid Earth*, **105**(B5), 11127-11152 (2000).

36. M. van Driel, L. Krischer, S. C. Stähler, K. Hosseini, T. Nissen-Meyer, Instaseis: instant global seismograms based on a broadband waveform database, *Solid Earth*, 6, 701-717, (2015)

37. J. Ritsema, E. Garnero, T. Lay, A strongly negative shear velocity gradient and lateral variability in the lowermost mantle beneath the Pacific. *Journal of Geophysical Research: Solid Eaerth*, 102(B9), 20395-20411. (1997).

38. L. V. D. Maaten, G. Hinton, Visualizing data using t-SNE. *Journal of Machine Learning Research*, **9**(Nov), 2579-2605 (2008).

39. Y. Rubner, C. Tomasi, L. J. Guibas, A metric for distributions with applications to image databases. *In Sixth International Conference on Computer Vision* (IEEE Cat. No. 98CH36271) (pp. 59-66). IEEE (1998).

40. J. M. Kolb, V. Lekic, Receiver function deconvolution using transdimensional hierarchical Bayesian inference. *Geophysical Journal International*, **193**(3), 1791-1735. (2014).

41. J. Tromp, D. Komatitsch, Q. Liu, Spectral-element and adjoint methods in seismology. *Communications in Computational Physics*, **3**(1):1–32. (2008).

42. A. M. Dziewonski, T.-A. Chou, J. H. Woodhouse, Determination of earthquake source parameters from waveform data for studies of global and regional seismicity, *J. Geophys. Res.*, **86**, 2825-2852. (1981).

43. J.W. Hernlund, C. Houser, On the statistical distribution of seismic velocities in Earth's deep mantle. *Earth and Planetary Science Letters*, **265**(3-4), pp.423-437. (2008).

44. F. A. Dahlen, S. H. Hung, G. Nolet, Fréchet kernels for finite-frequency traveltimes – I. Theory, *Geophysical Journal International*, **141**(1), pp. 157-174. (2000).

## Acknowledgments

We thank B. Romanowicz, A. To, S. Cottaar, N. Schmerr, and N. Guttenberg for insightful discussions that helped improve this manuscript. We acknowledge the thorough and thoughtful reviews from three anonymous reviewers, which also greatly improved the manuscript. **Funding:** This work is supported by Packard Foundation Fellowships to V.L. and B.M. and NSF-EAR1352214 grant to V.L. Development of the Sequencer was supported by the generosity of Eric and Wendy Schmidt by recommendation of the Schmidt Futures program. **Authors contributions:** D.K., V.L., and B.M. provided the intellectual framework, designed the research, and wrote the manuscript. D.K. performed the data analysis and computations and prepared the figures and technical details. B.M. and D.B. developed the Sequencer, while MT-P implemented the online Sequencer used for this research. All authors helped edit the manuscript. **Competing interests:** The authors declare no competing interests. **Data and materials availability**: Seismic data used in this manuscript are available through the IRIS Data Management Center (DMC). The source code of the Sequencer, together with implementation details, can be found at http://github.com/dalya/Sequencer/. An online version of the algorithm is available at http://sequencer.org.

## Supplementary Materials

Materials and Methods

Figures S1-S16

References (*36-44*)

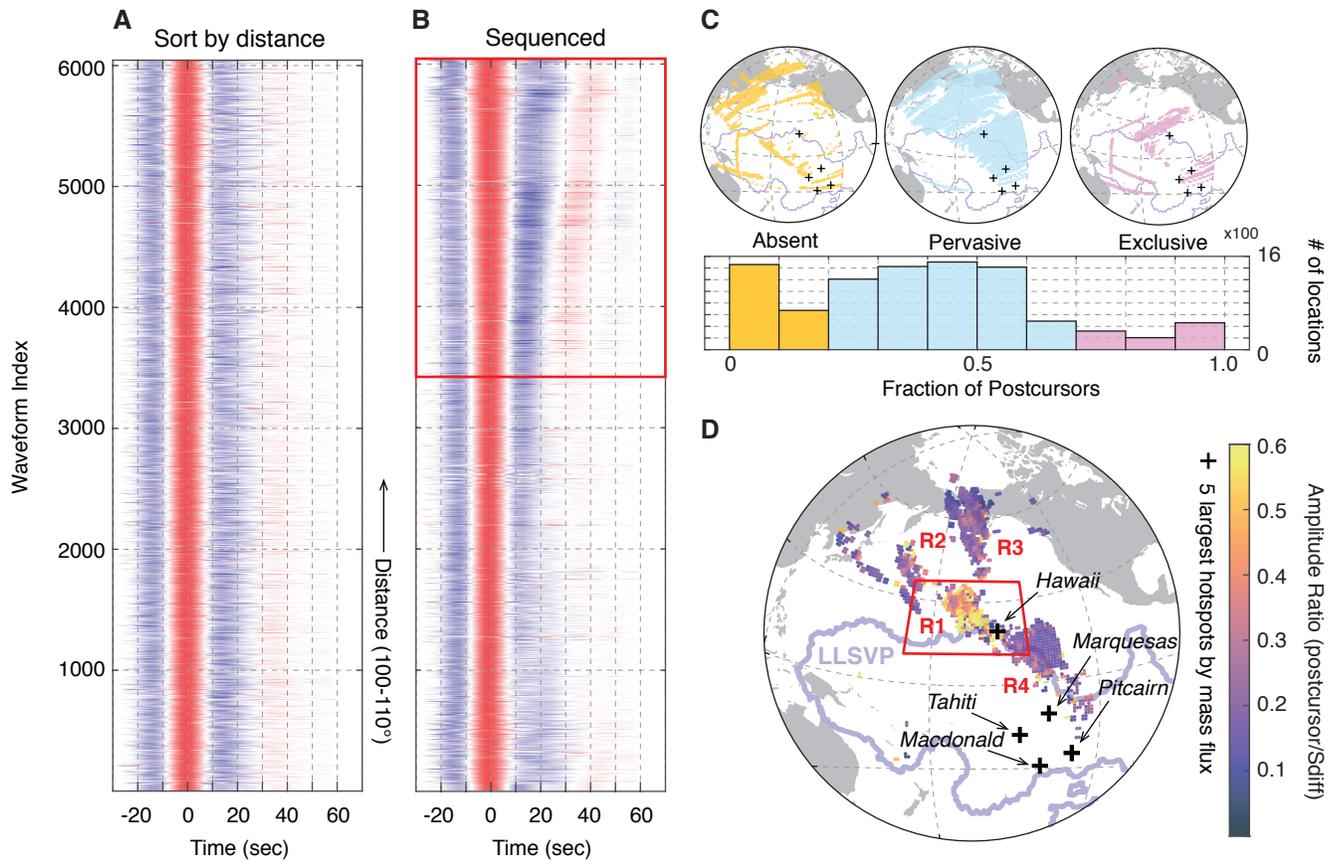

**Fig 1. Postcursors across the northern Pacific.** A comparison of the source deconvolved Sdiff waveforms of moment magnitude (Mw) > 6.5 deep earthquakes (depth > 200 km) between 100° and 110° epicentral distance sorted by (**A**) distance and (**B**) the Sequencer. Sequencer ordering enables the identification of a substantial (~40% of all waveforms) subpopulation of Sdiff postcursors [red box in (B)]. (**C**) Histogram of fraction of waveforms with postcursors in 1° radius bins (bottom) and the geographic locations (top) where postcursors are absent (peach), pervasive (cyan), and exclusive (pink). (**D**) Stacks of postcursor amplitude relative to main Sdiff arrival, also averaged in 1° bins. The geographic extent of the Pacific LLSVP (18) is shown (light-blue contour), as are the five largest hot spots by mass flux (black crosses) (35). The region with the strongest postcursors is outlined in red. Regions labeled R1 to R4 are discussed in the main text.

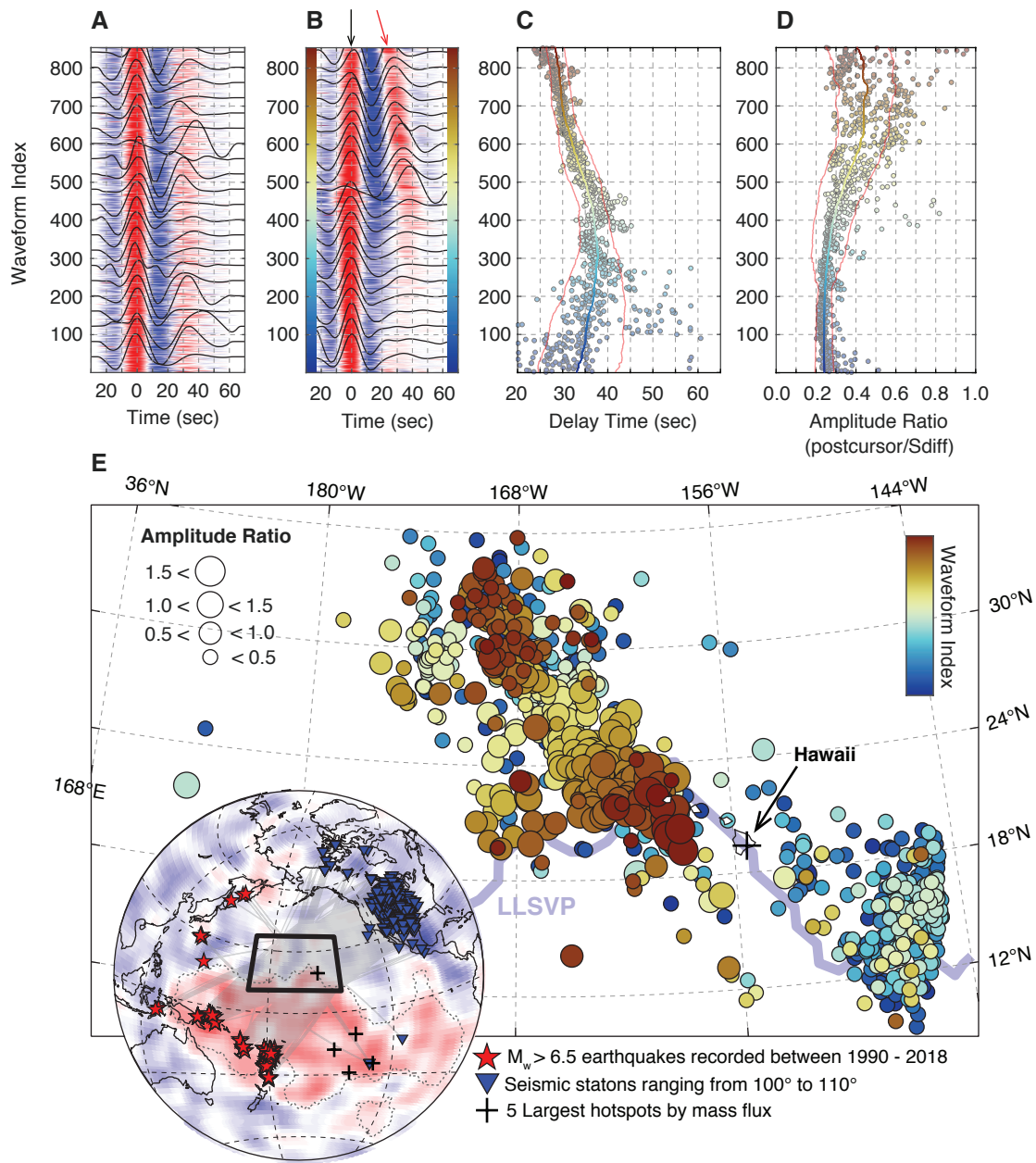

**Fig. 2. Postcursors in the Hawaii region.** Deconvolved Sdiff waveforms with turning points within 20° of Hawaii, sorted by (**A**) distance and (**B**) the Sequencer. Strong postcursors (red arrow) arrive after the main Sdiff phase (black arrow). Color bar represents sequence index and is used in later panels. The (**C**) delay time and (**D**) amplitude ratio of largest post-Sdiff arrivals, and (**E**) midpoints of their corresponding raypaths. The sequence reveals a coherent geographic pattern to the region northwest of Hawaii. Earthquakes and seismic stations are plotted as red stars and blue triangles (inset), respectively. The geographic extent of the Pacific LLSVP (18) is shown (purple contour), as are the five largest hot spots by mass flux (black crosses) (35).

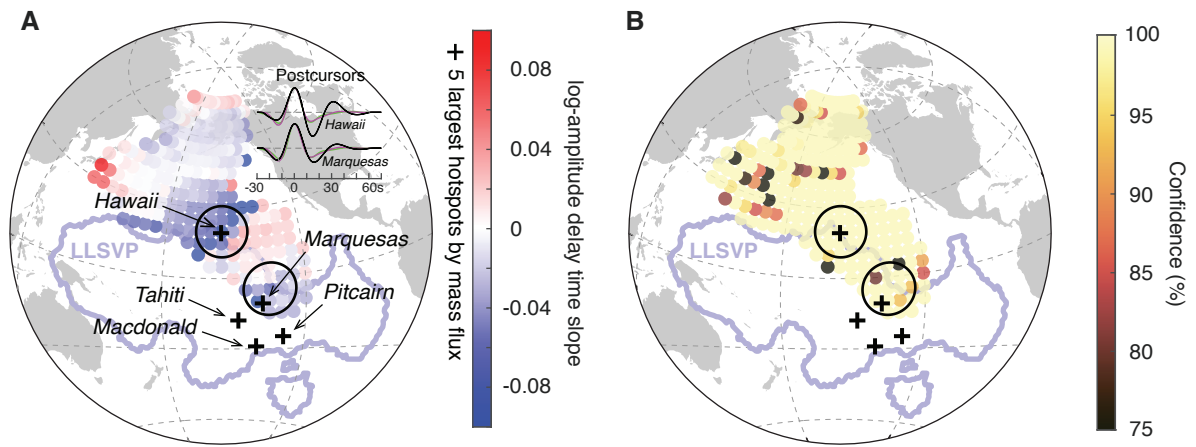

**Fig. 3. Detection of localized structures.** Map of (A) slope and (B) confidence range of the relationship between delay times and log amplitudes for postcursors identified in Fig. 1B and corresponding to raypaths that diffract within 5° of each location. Inset in (A) shows the average postcursor waveforms in Hawaii and the Marquesas (black), average of all nonpostcursor waveforms (green), and average of synthetic waveforms from PREM (magenta). Significance of these postcursor waveforms is discussed in the materials and methods (10) (fig. S10). Large negative value of the slope cluster around Hawaii and close to the Marquesas, indicating the presence of a localized source to the postcursors there. The geographic extent of the Pacific LLSVP (18) is shown (purple contour), as are the five largest hot spots by mass flux (black crosses) (35).

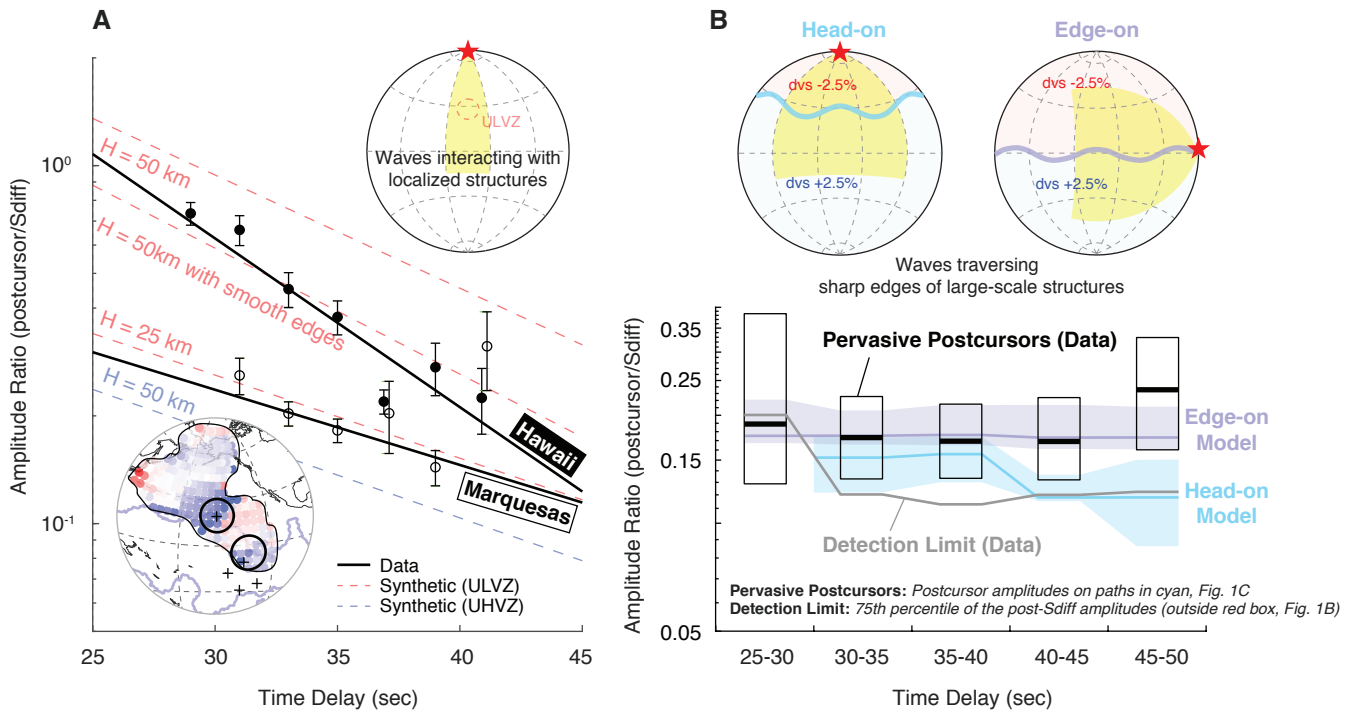

**Fig. 4. Comparison of observed postcursor delay time versus log amplitude trends with predictions for candidate lowermost mantle structures.** (A) Log amplitudes decrease strongly with delay time for postcursors near Hawaii (filled circles) and Marquesas (open circles) hot spots, in agreement with predictions (dashed lines) for three cylindrical ULVZ models (red). The prediction for an ultrahivgh-velocity zone (UHVZ) model is plotted in blue. Upper inset shows illumination geometry of synthetic waveforms calculated for the cylindrical ULVZ models. Lower inset shows Fig. 3A and the diffraction coverage of the dataset (black outline). (B) For pervasive postcursors, log amplitude is observed to be constant with delay time (black boxplots) and agrees with the postcursors predicted for waves traversing a sharp LLSVP boundary (5% Vs contrast) edge-on (lavender). Head-on geometries (light blue) underpredict the amplitudes. Detection limit is defined by the 75th percentile of the post-Sdiff amplitude in nonpostcursor waveforms (outside the red box in Fig. 1B).